\documentclass[english,prd,superscriptaddress,nofootinbib,preprintnumbers,twocolumn,showpacs]{revtex4}
\usepackage[utf8]{inputenc}
\usepackage[english]{babel}
\usepackage{amsmath}
\usepackage{amsfonts}
\usepackage{amssymb}
\usepackage{epsfig}
\usepackage{graphics,psfrag,rotating}
\usepackage{graphicx}
\usepackage{dcolumn}
\usepackage{bm}
\usepackage{epstopdf}
\usepackage{color}
\usepackage[usenames,dvipsnames,svgnames]{xcolor}
\usepackage[colorlinks=true,
            linkcolor=red,
            urlcolor=gray,
            citecolor=blue]{hyperref}
  \usepackage{hyperref}
\usepackage[T1]{fontenc}
\usepackage{multirow}
\usepackage{float}

\usepackage{subfigure}

\usepackage{enumitem}

\def\nab{\nabla}

\def\prjnab{ \widetilde{\nabla}}

\def\prjnab{ \widetilde{\nabla}}

\def\nab{\nabla}
\def\3nab{\tilde{\nabla}}

\def\c{\mbox{curl}}

\def\tl{\tilde}
\newcommand{\sfrac}[2]{{\textstyle{#1\over#2}}}
\def\be {\begin{equation}}
\def\ee {\end{equation}}
\def\ba {\begin{align}}
\def\ea {\end{align}}

\def\bc {\begin{center}}
\def\ec {\end{center}}
\def\case#1/#2{\frac{#1}{#2}}

\newcommand{\bver}{\begin{verbatim}}
\newcommand{\ever}{\end{verbatim}}

\newcommand{\nb}{\nabla}

\newcommand{\bea}{\begin{eqnarray}}
\newcommand{\eea}{\end{eqnarray}}
\newcommand{\beaa}{\begin{eqnarray*}}
\newcommand{\eeaa}{\end{eqnarray*}}

\def\case#1/#2{\textstyle\frac{#1}{#2}}

\begin{document}
\title{Constraining the gravitational action with CMB tensor anisotropies} 
\author{Mohamed Abdelwahab}
\affiliation{Astrophysics Cosmology \& Gravity Center, and Department of Mathematics \& Applied Mathematics,
University of Cape Town, 7701 Rondebosch, South Africa.}
\author{\'Alvaro\,de la Cruz-Dombriz}
\affiliation{Departamento de F\'{\i}sica Te\'orica I, Ciudad Universitaria, Universidad Complutense de Madrid, E-28040 Madrid, Spain.}
\author{Peter Dunsby}
\affiliation{Astrophysics Cosmology \& Gravity Center, and Department of Mathematics \& Applied Mathematics,
University of Cape Town, 7701 Rondebosch, South Africa.}
\affiliation{South African Astronomical Observatory,  Observatory 7925, Cape Town, South Africa.}
\author{Bishop Mongwane}
\affiliation{Astrophysics Cosmology \& Gravity Center, and Department of Mathematics \& Applied Mathematics,
University of Cape Town, 7701 Rondebosch, South Africa.}

\pacs{04.30.-w, 04.50.Kd, 98.80.-k}

\begin{abstract}
We present a complete analysis of the imprint of tensor anisotropies on the Cosmic Microwave Background for a class of $f(R)$ gravity theories within the PPF-CAMB framework.  We derive the equations, both for the cosmological background and gravitational wave perturbations, required to obtain the standard temperature and polarisation power spectra, taking care to include all effects which arise from $f(R)$ modifications of both the background and the perturbation equations. For $R^{n}$ gravity, we show that for $n\neq 2$, the initial conditions in the radiation dominated era are the same as those found in General Relativity.  We also find that by doing simulations which involve either modifying the background evolution, while keeping the perturbation equations fixed, or fixing the background to be the $\Lambda$CDM model and modifying the perturbation equations,  the dominant contribution to deviations from General Relativity in the temperature and polarisation spectra can be attributed to modifications of the background. This demonstrates the importance of using the correct background model in perturbative studies of $f(R)$ gravity. Finally an enhancement in the $B$ modes power spectra is observed which may allow for 
lower inflationary energy scales. 
\end{abstract}
\maketitle
\section{Introduction}
Modified theories of gravity have become one of the most popular candidates for explaining the current accelerated expansion of the Universe. As it is well-known,  General Relativity (GR) in its standard form (without a cosmological constant) can not explain such cosmological speed-up, without the introduction of extra terms in the gravitational Lagrangian (for  reviews on modified theories of gravity, see Refs.~\cite{review,book}) or exotic fluid components (see Refs.~\cite{Ref1,Ref2,Ref3}).
Geometrical modifications of the standard gravitational Lagrangian usually include a wide number of curvature invariants ($R$, $R_{\mu\nu}R^{\mu\nu}, R_{\mu\lambda\nu\sigma}R^{\mu\lambda\nu\sigma}$,..)  \cite{Gauss-Bonnet} that might imply non-minimal couplings between matter and geometry (see for instance \cite{Non-minimal} and references therein).
The simplest and most studied modification of GR is one where the Hilbert-Einstein action is generalised to a arbitrary function of the Ricci scalar $R$ - these are the so-called $f(R)$ theories of gravity \cite{Ref5}-\cite{Ref6}. These theories are able to mimic the behaviour of the cosmological constant (see for instance Ref.~\cite{gr-qc/0607118}) and reproduce the entire background cosmological history (see Ref.~\cite{unification}). Despite these successes, $f(R)$ theories have their own shortcomings \cite{Ref6} and specific models have to pass rigorous theoretical and observational scrutiny before they can be accepted as viable alternatives to the Concordance Model \cite{Viable}. 

In the last few years, modified theories of gravity have been shown not only able to mimic the dark energy epoch, but also seem capable of providing a description of the inflationary era \cite{unification}.  Unfortunately, the use of cosmological observations, such as supernovae type Ia or Baryon Acoustic Oscillations (BAO) among others, which depend solely upon the expansion history of the Universe is not enough to only determine the nature and the origin of dark energy, due to the fact that identical evolutions for the cosmological background can be explained by a diverse number of theories. This is the so-called {\it degeneracy problem}. In order to test the validity of extended theories of gravity, it is therefore necessary to correctly describe the growth rate and matter power spectra as obtained from  scalar perturbations (see Refs.~\cite{Perturbations} for analytical techniques and Refs. \cite{Perturbations-numerics} for numerical parameterisations with several codes) and the existence and stability of astrophysical objects such as black holes \cite{BH}.  Another powerful tool to discriminate between competitive theories of gravity consists in studying both the temperature and the polarisation modes in the Cosmic Microwave Background (CMB) power spectra. Since those spectra are made of contributions from both scalar and tensor perturbations, potential signatures from CMB spectra obtained from extended theories are expected.

The study of the effect of tensor (gravitational wave) perturbations on the CMB in modified gravity has not received much attention in comparison with their scalar counterpart, where the focus has been on the study of the evolution of the density contrast in these theories \cite{Perturbations}.  This is due to the difficulties modifying the standard CMB codes available for General Relativity, such as CAMB \cite{CAMB_Lewis}  or CMB-Easy \cite{CMB-Easy} (originally based on CMBFast \cite{CMB-Fast, CMB-Fast-link}).

A number of different attempts to study CMB tensor anisotropies have been made for several modified gravity scenarios. For instance,  the contribution made by cosmic strings in an Abelian Higgs model to the temperature power spectrum of the CMB was studied in \cite{Bevis:2006mj}.  Calculations in this paper were performed in a modified version of CMB-Easy and were able to constrain the string tension when normalised with the WMAP data available at that time. A number of predictions in the CMB spectrum for this model were presented in \cite{Bevis_predictions}. Also, in the last few years, some attention has been devoted to the study of CMB tensor perturbations in brane-world theories.  For instance,  in \cite{Battye:2003ks} the evolution of cosmological tensor perturbations in the Randall-Sundrum II model was discussed. In the near-brane limit, the separation of the wave equations becomes possible, making the study of the evolution of perturbations feasible. Other work \cite{Challinor} focused on extending the CAMB code \cite{CAMB_Lewis} to compute CMB tensor anisotropies generated in a generalised $1+4$ Randall-Sundrum II braneworld model. 

Finally, with regard to $f(R)$ fourth-order gravity theories, the only existing attempt to encapsulate the main features of tensor perturbations was presented in \cite{Zhong:2010ae}. Here the authors analysed tensor perturbations of flat thick domain wall branes in  $f(R)$ gravity. They showed that under the transverse and traceless gauge, the metric perturbations decoupled from perturbations of the background scalar field, which generates the brane. 

The leitmotiv of this investigation is to address for the first time in literature a detailed analysis of the effect of tensor perturbations on the CMB in a class of metric $f(R)$ gravity theories, when studied in the so-called Jordan frame.  This provides a further way of determining whether such models are viable, particularly in situations where the background evolution is close to that of the standard $\Lambda$CDM model.

The paper is organised as follows:  in Section \ref{Generalities} we present the general features of $f(R)$ theories and introduce the different fluids that play a role in the cosmological evolution. In Section \ref{Dynamics of f(R) theories} we introduce the background and tensor perturbation dynamics for these theories, paying special attention to the initial conditions set in the radiation dominated era. Section \ref{Rn_models} is then devoted to illustrating the formalism for a simple class of one-parameter $f(R)$ models by obtaining both the cosmological evolution and the CMB features resulting from tensor perturbations.  Section \ref{Results} then summarises the main results we obtain once the simulations were implemented. 

We conclude by giving our conclusions in Section \ref{Conclusions}.  At the end of the paper we have included an Appendix with specific details on the relevant CAMB modifications.
\section{Fourth order gravity} 
\label{Generalities}
The most general action for fourth-order gravity \cite{reviw1, reviw2, reviw3} can be written as a function of the Ricci scalar only.
\begin{equation}
\label{lagr f(R)}
\mathcal{A}\,=\,\frac{1}{2}\int \text{d}^4 x \sqrt{-g}\left[ f(R)+{\cal L}_{m}\right]\,,
\end{equation}
where $\mathcal{L}_m$ represents the matter contribution and geometrised units $8\pi G\equiv 1$ are used. The variation of this action with respect to the metric leads to the generalised Einstein equations in the metric formalism: 
\begin{equation}
\label{eq:einstScTn}
f_R G_{ab} =T^{m}_{ab}+ \case{1}/{2} \left(f-Rf_R\right) g_{ab} 
+ \nabla_{b}\nabla_{a}f_R- g_{ab}\nabla_{c}\nabla^{c}f_R\,,
\end{equation}
where $f\equiv f(R)$, $f_R\equiv {\rm d}f/{\rm d}R$ (analogous notation for higher derivatives in the following) and
\begin{equation}
T^{m}_{ab}\equiv\frac{2}{\sqrt{-g}}\frac{\delta
(\sqrt{-g}\mathcal{L}_{m})}{\delta g_{ab}} 
\end{equation}
represents the energy-momentum tensor of standard matter. These equations obviously reduce to the standard Einstein field equations when $f(R)=R$. It is crucial for our purposes to be able to rewrite expression \eqref{eq:einstScTn} in the form,
\begin{equation}
\label{eq:einstScTneff}
 R_{ab}-\case{1}/{2}R\, g_{ab}\,=\, T^{total}_{ab}\,=\,\tilde{T}_{ab}^{m}+T^{R}_{ab}\,, 
 \end{equation}
where $\tilde{T}_{ab}^{m}$ and  $T_{ab}^{R}$ are defined as follows
\begin{eqnarray}
\label{eq:TenergymomentuEff}
\tilde{T}_{ab}^{m}&=&\frac{T_{ab}^{m}}{f_R}\,,\nonumber\\
T_{ab}^{R}&=&\frac{1}{f_R}\left[\case{1}/{2} \left(f-R f_R\right)g_{ab} +\nab_b\nab_a f- g_{ab}\nab_c\nab^cf\right]\;, \nonumber\\
\label{eq:semt}
\end{eqnarray}
that represent respectively two effective {\em fluids}:
the {\em effective matter fluid} (associated with $\tilde{T}_{ab}^{m}$) and
the {\em curvature fluid} (associated with $T^{R}_{ab}$). The twice contracted Bianchi identities for these fluids are:
\begin{eqnarray}
\label{BI}
\tilde{T}_{ab}^{m;b}&=&\frac{{T}_{ab}^{m;b}}{f_R}-\frac{f_{RR}}{f_R^{2}}\,{T}_{ab}^{m}\,R^{;b}\,,\nonumber\\
T^{R;b}_{ab}&=&\frac{f_{RR}}{f_R^{2}}\,{T}_{ab}^{m;b}\,R^{;b}\,.
\end{eqnarray}
where the semicolon denotes a covariant derivative. It is clear from these identities that the total energy-momentum tensor ${T}_{ab}^{total}$ as defined in \eqref{eq:einstScTneff} will always be divergence-less provided that ${T}_{ab}^{m}$ is divergence-less. In the matter frame $u^a$ the total energy momentum tensor can be decomposed as follows:
\begin{eqnarray}
\label{EMT}
T_{ab}^{total}=\mu\,u_a\,u_b+p\,h_{ab}+q_a\,u_b+q_b\,u_a+\pi_{ab}\,,
\end{eqnarray}
where $\mu$ is the total energy density and $p$ is the total isotropic pressure of the fluid, $q_a$ represents the total energy flux, $\pi_{ab}$ is the total anisotropic pressure. In terms of the two effective fluids \eqref{BI}, these effective thermodynamical quantities can be written as,
\begin{eqnarray}
\label{thermodynamics_quantities}
\mu &=&\frac{\mu_m}{f_R}+\mu^R\,,~~ p=\frac{p_m}{f_R}+p^R\,, \nonumber\\
q_a&=&\frac{q_a^m}{f_R}+q_a^R\,, ~~\pi_{ab}=\frac{\pi_{ab}^m}{f_R}+\pi_{ab}^R\,.
\end{eqnarray}
In order to extract those thermodynamical quantities from the field equations, we use the standard technique of projecting onto orthogonal surfaces to the $4$-velocity of the fluid flow $u^a$ using the projection tensor $h_{ab}\equiv g_{ab}+u_{a}u_{b}$. We refer the reader to \cite{Amare} (and references therein) for further details on the 1+3 covariant approach. The geometry of the flow lines is determined by  the kinematics of $u^{a}$, which allows us to relate important kinematic quantities via
\be
\nb_{b}u_{a}=-u_{b}A_{a}+\sfrac{1}{3}\Theta h_{ab}+\sigma_{ab}+\omega_{ab}\;,
\ee
where the right-hand side of this equation contains the acceleration $A^{a}$ of the standard matter, expansion $\Theta$, shear $\sigma_{ab}$ and vorticity $\omega_{ab}$, which will be used in the following sections.

The effective thermodynamical quantities (\ref{thermodynamics_quantities}) for the curvature fluid then become \cite{Amare}:
\begin{eqnarray}
\label{ETQ}
&&\mu^{R}\,=\,\frac{1}{f_{R}}\left[\frac{R f_{R}-f}{2}-\Theta
f_{RR}\dot{R}+f_{RR}\tilde{\nabla}^2{R}\right]\;,\label{muR}\nonumber\\
&&p^{R}\,=\,\frac{1}{f_{R}}\left[\frac{f-R
f_{R}}{2}+f_{RR}\ddot{R}+f_{3R}\dot{R}^2+\case{2}/{3}\Theta
f_{RR}\dot{R}\right]\;,\label{pR}\nonumber\\
&&\nonumber\\
&&q^{R}_a\,=\,-\frac{1}{f_{R}}\left[f_{3R}\dot{R}\prjnab_{a}R+f_{RR}\prjnab_{a}\dot{R}-\frac{1}{3}\Theta f_{RR}
\prjnab_{a}R\right]\;,\nonumber\\
&&\nonumber\\\
&&\pi^{R}_{ab}\,=\,\frac{1}{f_{R}}\left[f_{RR}\prjnab_{\langle
a}\prjnab_{b\rangle}R-\sigma_{a
b}f_{RR}\dot{R}\right]\,,\label{piR}
\end{eqnarray}
where  $\dot{R}=u_{a}\nb^{a}R$
and $\tl\nb_{a}=h^{b}{}_{a}\nb_{b}$ denote the covariant convective derivative on the Ricci scalar $R$ and the spatially totally projected covariant derivative operator orthogonal to $u^{a}$ respectively. Angular brackets denotes the projected, trace-free, symmetric part of a tensorial quantity \footnote{For instance 
\begin{eqnarray}
\label{vector}
W_{\langle ab\rangle}\equiv \left[h_{(a}{}^{c}{}h_{b)}{}^{d}-\sfrac{1}{3}h^{cd}h_{ab} \right]W_{cd}\,, \nonumber
\end{eqnarray}
provides the projected, trace-free part of a tensor $W_{ab}$.
}.
\section{Tensor perturbations}
\label{Dynamics of f(R) theories}
For a Friedmann-Lema\^{i}tre-Robertson-Walker (FLRW) background with vanishing 3-curvature and a barotropic perfect fluid source with equation of state $p=\omega\rho$  as the standard matter source, the independent field equations for general $f(R)$ gravity can be written as
\begin{eqnarray}
&&\Theta^2 =3\mu\,,
\nonumber\\
&&\dot{\Theta}=-\case{1}/{3}\Theta^2-\case{1}/{2}(\mu+3 p)\,, 
\nonumber\\
&&\dot{\mu}_m=-\Theta\,(\mu_m+p_m)\,,
\label{ConE}
\end{eqnarray}
known as the Friedmann, Raychaudhuri and energy conservation equations respectively.
The linearisation of the exact propagation and constraint equations around this background for a pure tensor perturbations then leads to the system \cite{per}:
\begin{eqnarray}
\label{eqsigma}
&&\dot{\sigma} _{ab}+ \case{2}/{3}\,\Theta \,\sigma_{ab}+ E_{ab} - \case{1}/{2} {\pi}_{ab}=0\,,
\nonumber\\ 
&&\dot{H}_{ab} + H_{ab}\,\Theta+(\c\,E)_{ab} - \case{1}/{2}(\c\,{\pi})_{ab} =0\,,\nonumber\\
&&\3nab_{b}H^{ab}=0\;,~ \3nab_{b}E^{ab}=0\;,~ H_{ab}=(\c\,\sigma)_{ab}\,,
\end{eqnarray}
together with the linearised conservation equations,
\begin{eqnarray}
\label{CE}
&&\dot{\mu}^m=-\Theta\,\mu^m\,(1+\omega)\;,\nonumber\\
&&\dot{\mu}^R=-\Theta\,\mu^R(1+\omega_{eff})\,,
\end{eqnarray}
that are obtained from the Bianchi identities contracted twice, where
 \begin{eqnarray}\label{eff}
\omega_{eff}\,\equiv\,\frac{p^R}{\mu^R}-\frac{\dot{R}\,f_{RR}\,\mu^m}{f_{R}^2}
\end{eqnarray}
is the effective dark energy equation of state and $w$ is the equation of state of standard matter. Taking the time derivative of equations (\ref{eqsigma}) and using the standard tensor harmonic decomposition \cite{per}:
\begin{eqnarray}
\sigma_{ab} &=& \sum_{k} \frac{k}{a} \left[\sigma_k Q_{ab}^{(k)} + \bar{\sigma}_k \bar{Q}_{ab}^{(k)}\right],  \nonumber\\
\pi_{ab} &=& \rho \sum_{k} \left[ \pi_{k} Q_{ab}^{(k)} + \bar{\pi}_k \bar{Q}_{ab}^{(k)} \right]\,,
\end{eqnarray}
we obtain a second order equation for the tensor modes $\sigma_k$:
\begin{eqnarray}\label{swe}
\ddot{\sigma}_k+\Theta\,\dot{\sigma}_k+\Big[\frac{k^2}{a^2}-\frac{1}{3}(\mu+3\,p)\Big]\,\sigma_k=\nonumber\\
\frac{a}{k}\,\Big[\mu\, \dot{\pi}_{k}-\frac{1}{3}(\mu+3\,p)\,\Theta\,\pi_k\Big]\,.
\end{eqnarray}
Once the form of the anisotropic pressure has been determined, the above equations can be solved to give the evolution of tensor perturbations. From equation \eqref{piR} and for pure tensor modes, we obtain:
\begin{eqnarray}
\label{pigc}
\pi_k^{R}=-\frac{k}{a\, \mu}\frac{f_{RR}\,\dot{R}}{f_{R}}\,\sigma_k\,. 
\end{eqnarray}
It is clear that since $\pi_k$ is proportional to $\sigma_k$, 
that the tensor perturbation equations will be second order  for $f(R)$ theories, unlike their scalar counterparts \cite{Perturbations}, where the involved equations are in general fourth order.
\subsection{The initial conditions}
In the radiation dominated era, the anisotropic stress $\pi$ is dominated by the radiation fluid contribution. Therefore, in this epoch $\pi =\pi^\gamma$ and consequently, equation (\ref{swe}) reduces to
\begin{eqnarray}
\label{shd}
&&\ddot \sigma_k + A \dot \sigma_k + B \sigma_k \nonumber\\
&&=\frac{a}{k} \frac{\mu}{f_{R}}
 \left\{ \dot \pi^\gamma_k - \left[H\left(1+3\omega\right) +
\frac{f_{RR}}{f_{R}} \right]\pi^\gamma_k \right\}\;, \end{eqnarray}
where the quantities $A$ and $B$ are given by
\begin{eqnarray}
&&A \equiv 3H + \frac{f_{RR}}{f_{R}} \dot R\,,\nonumber\\
&&B\equiv\frac{k^2}{a^2} +  \frac{2\ddot a}{a} + \dot R^2 \left[ \frac{f_{3R}}{f_{R}} -\left(
\frac{f_{RR}}{f_{R}} \right) ^2 \right]\nonumber\\
&&\;\;\;\;+\;\,\frac{f_{RR}}{f_{R}}\left(\ddot R+H\dot R\right) \,.\label{B}
\end{eqnarray}
Thanks to the homogeneity of the early universe, equation (\ref{shd}) can be further simplified by assuming that the radiation anisotropic stress vanishes. This gives:
\begin{eqnarray}
\label{se}
\sigma_k''+ \left( a A - \mathcal{H} \right) \sigma_k' + a^2 B \sigma_k =0\,,
\end{eqnarray}
where prime denotes derivative with respect to conformal time $\tau$ and $\mathcal{H}\equiv a'/ a = aH$. 
After performing the change $u_k =a^m \sigma_k$, where $m$ in general a function of time, equation (\ref{se}) reads
\begin{eqnarray}\label{IC3}
&&u_k''  + 
\left[aA -\mathcal{H}\left(1+2m\right) -2m' \log{a} \right] 
u_k'\nonumber\\
&&\;+ \left\{a^2 B - m \mathcal{H}' -2\mathcal{H} m' +
\left[-aA+\mathcal{H}(1+m)\right.\right.\nonumber\\
&&\left.\left.+\,m' \log{a}\right]
\left(m\mathcal{H}+m' \log{a}  \right)-m'' \log{a}
\right\}  u_k =0\,.\nonumber\\ &&
\end{eqnarray}
At this stage, let's remark that in the next section we will show that the exponent $m$ is constant for the case of $R^n$ models in the radiation dominated epoch. 
\section{$R^n$ gravity}
\label{Rn_models}
\label{Rn_models}
In order to illustrate the formalism described in the previous sections, we now consider a one parameter class of $f(R)$ theories: $f(R)=R^n$ and study the background and tensor perturbations evolution for these models.
\subsection{Evolution equations for the background}
%
Let's define the following set of dynamical variables for determining the expansion history for $R^n$ gravity. Following \cite{Dyn}, we introduce the dimensionless variables:
\begin{eqnarray}
\label{DV}
&&x=\dfrac{\dot{R} (n-1)}{H R}\;,~y=\dfrac{R (1-n)}{6 n H^2}\;,\nonumber\\
&&\Omega_d=\dfrac{\mu_d}{3 H^2 n R^{n-1}}\;,~\Omega_r =\dfrac{\mu_r}{3 H^2 n R^{n-1}}\;,
\end{eqnarray}
where $\mu_{d}$ and $\mu_{r}$ are the dust and radiation densities respectively. In terms of these variables, the Friedmann equation in (\ref{ConE}) takes the simple form\,,
\begin{eqnarray}\label{const}
1+x+y-\Omega_d-\Omega_r=0\,.
\end{eqnarray}
At this stage, an autonomous system, which is equivalent to cosmological equations (\ref{ConE}) can be derived by differentiating the dynamical variables defined in (\ref{DV}):
\begin{eqnarray}
\label{DS2}\nonumber
a\,\frac{\text{d}x}{\text{d}a}&=&-x-x^2+\dfrac{(4-2n+nx)y}{n-1}+\Omega_d\,,\\ \nonumber
a\,\frac{\text{d}y}{\text{d}a}&=&\left[4+\dfrac{(x+2ny)}{n-1}\right]\,y\,,\nonumber \\
a\,\frac{\text{d}\Omega_d}{\text{d}a}  &=&\left[ 1-x+\dfrac{2ny}{n-1}\right] \Omega_d\,, \\ \nonumber
a\,\frac{\text{d}\Omega_r}{\text{d}a} &=&\left[-x+\dfrac{2ny}{n-1}\right] \Omega_r\,,\\ \nonumber
\end{eqnarray} 
The constraint equation (\ref{const}) can be used to reduce the dimensionality of the system above. The evolution equation of Hubble parameter $H$ is given in terms of the dynamical variables (\ref{DV}) by\,,
\begin{eqnarray}
\label{H}
a\,\frac{\text{d}H}{\text{d}a}&=&-H\left(2+\dfrac{ny}{n-1}\right)\,.
\end{eqnarray} 
Furthermore, the deceleration parameter $q$ can be determined directly from the definition of the variable $y$ as follows:
\be
q=\frac{ny}{(n-1)}+1\;.
\ee
The fixed points of the system (\ref{DS2}) are shown in Table \ref{Table} according to the results in \cite{Dyn}. In order to study the stability of these fixed points we use the well-known techniques, which involve linearizing the dynamical equations around the equilibrium points and then finding the eigenvalues of the linearisation matrix -- the Jacobian --  at these points. 
There are three interesting points in the phase space of $R^n$-gravity models:  the points J and G correspond respectively to transient radiation and matter dominated decelerated power-law expansions and the point B represents an accelerated expansion phase for particular values of the parameter $n$. In \cite{Dyn} it was also shown that phases represented by these three points provide cosmological evolutions with positive energy density if $n$ lies in the range $1.36<n<1.5$.

In fact, a large number of orbits connecting these three points can be found. Since we are interested in a background evolution that is similar to $\Lambda{\rm CDM}$, we used a numerical procedure to single out the orbit which gives the best fit to $\Lambda{\rm CDM}$ evolution. In order to illustrate this procedure, we chose values around $n \simeq 1.28$ which provided Hubble parameters today sufficiently close to $\Lambda$CDM and reasonable fits for BAO and large-scale structure data \cite{Amare_PRD2013}. It should be pointed out that these values of $n$ are not able to  provide acceleration during the present epoch.
 
\begin{table}
\caption{ Coordinates of the fixed points for$R^{n}$-gravity according to \cite{Dyn}.
Points J and G represent transient decelerated power-law expansion phases for radiation and and matter epochs respectively, whereas point B corresponds to a late-time accelerated expansion phase for certain values of $n$. Trajectories in the phase space connecting these three points can therefore represent realistic cosmological expansion histories.\\}
\begin{tabular}{lllll}
\hline
Point & Coordinates [$x$, $y$, $\Omega_d$, $\Omega_r$] & \\
\hline
A & $\left[-1,0,0,0\right]$    &  \\
B & $\left[\dfrac{4-2n}{1-2n},\dfrac{5-4n}{2n-1},0,0\right]$&  \\
C & $\left[-2,0,0,-2\right] $ &  \\
D & $\left[0,0,0,0\right]$ &  \\
E & $\left[2-2n,-2(-1+n)^2,0,0\right]$ &  \\
F & $\left[0,0,0,1\right]$ &  \\
G & $\left[-3+\dfrac{3}{n},\dfrac{-3+(7-4n)n}{2n^2},\dfrac{-3+(13-8n)n}{2n^2},0\right]$&  \\
H & $\left[-1,0,-1,0\right]$ &  \\
I & $\left[1,0,2,0\right]$ &  \\
J & $\left[-4+\dfrac{4}{n},-\dfrac{2(n-1)^2}{n^2},0,-5+\dfrac{8n-2}{n^2}\right]$&  \\
\hline
 \label{Table}
\end{tabular}
\end{table}
\subsection{Tensor Perturbations}
For $f(R)=R^n$ models, the scale factor in the radiation dominated era satisfies $a(\tau)=\tau^{\frac{n}{2+n}}$ \cite{Francaviglia}. Consequently, the coefficient in the damping term  in (\ref{IC3}) can be cancelled 
provided that the choice $m=\frac{Aa - \mathcal{H}}{2 \mathcal{H}}=\frac{2-n}{n}$ is made, where we have used (\ref{B}). 
In this case $m$ is constant and allows us to simplify (\ref{IC3}) as follows \cite{CMB-ERE2011}:
\begin{eqnarray}
\label{IC2}
u_k'' + \left( k^2-2\,\tau^{-2}\right) u_k =0\,,
\end{eqnarray}
where we have avoided the pathological case $n\neq 2$.
The result in \eqref{IC2} is exactly the same as the one in GR for tensor perturbations obtained in \cite{Lewis}. On the other hand,
for $R^n$ models equation (\ref{pigc}) becomes
\begin{eqnarray}
\label{pi}
\pi_k^R=-\frac{k}{a^2}\frac{(n-1)}{\mu\,R} R' \,\sigma_k\;,
\end{eqnarray}   
whose importance was precisely stressed after \eqref{pigc}. 
\section{Tensor power spectra}
\label{Results}
In the latest version of CAMB \cite{CAMBnew},
the possibility of using an external data file as input for the equation of state parameter of the effective dark energy component exists. Within $f(R)$ theories of gravity, the curvature fluid \eqref{eq:semt} is expected to play the role of dark energy. In this case, equation \eqref{eff} is used to generate the required data file for the effective equation of state parameter $w_{eff}$ of the curvature fluid. By supplying the ppf-CAMB code with this data file and using the effective matter density $\mu_{eff}\equiv \mu/f_{R}$, together with equation \eqref{pi}, we were able to generate the correct background evolution in CAMB (see Appendix \ref{sec:appendix} for further details on the relevant CAMB modifications). This procedure was usually missing in previous investigations which, for the sake of simplicity, simply assumed a $\Lambda$CDM background when studying CMB anisotropies of modified gravity theories.

\smallskip
Once the concomitant modifications in the background and perturbation equations have been implemented within the CAMB package, we consider different values of $n$ to study the signatures of modified gravity in the CMB tensor power spectrum. In particular, we consider the values $n=\{1.27, 1.29, 1.30, 1.31\}$. This choice of $n$ values is motivated by the fact that the best fit $n$ value for $R^n$ models to BAO data was obtained for $n\simeq 1.29$ \cite{Amare_PRD2013}.
In general, we expect distinct features of modified gravity depending on the particular value of $n$ on both the $c_{\ell}^{{TT}}$, $c_{\ell}^{{EE}}$ and $c_{\ell}^{{BB}}$ coefficients to be visible when compared to the standard $\Lambda$CDM results.

\begin{figure}[htb!]
\includegraphics[width=0.5\textwidth]{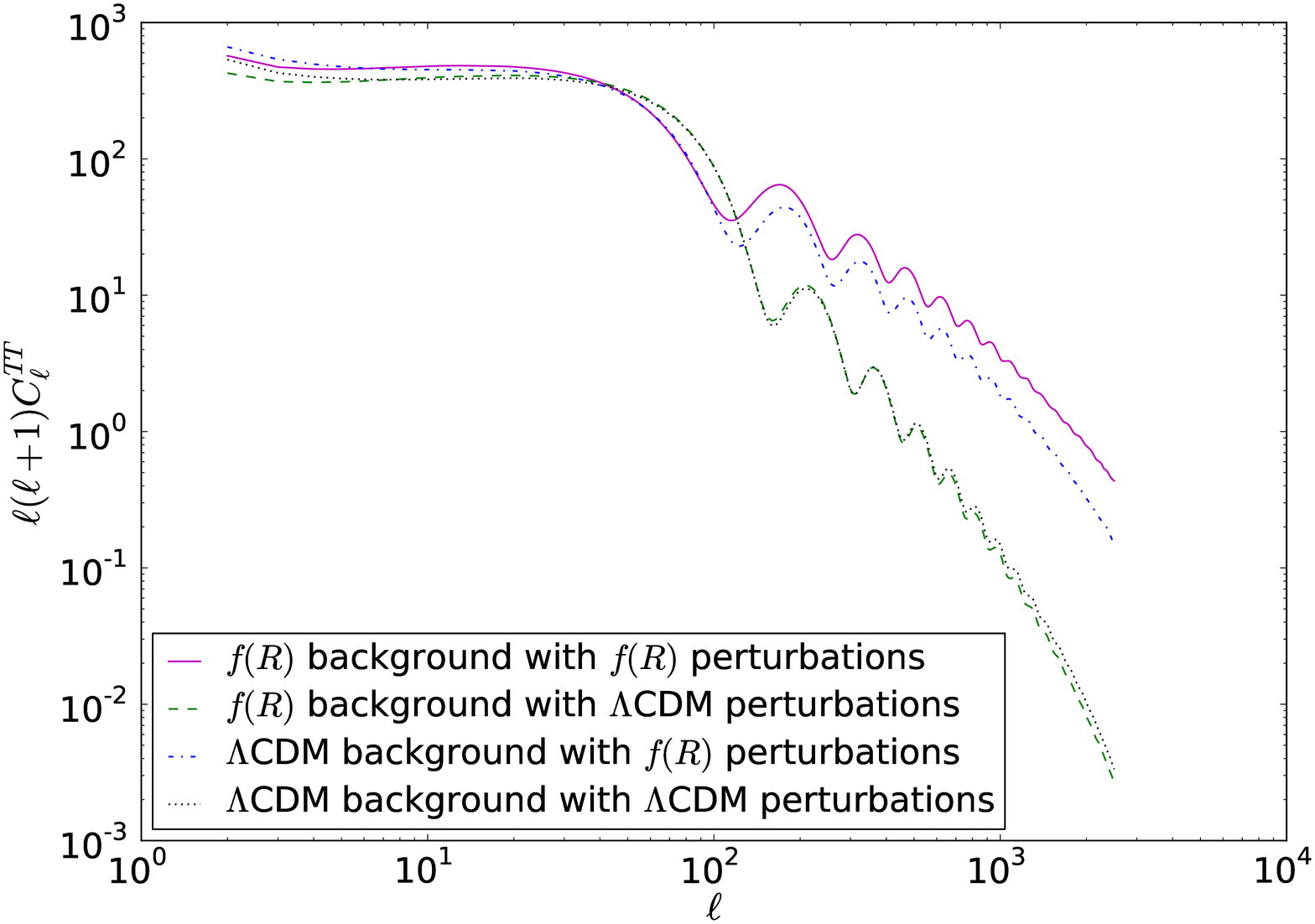}
\\
\includegraphics[width=0.5\textwidth]{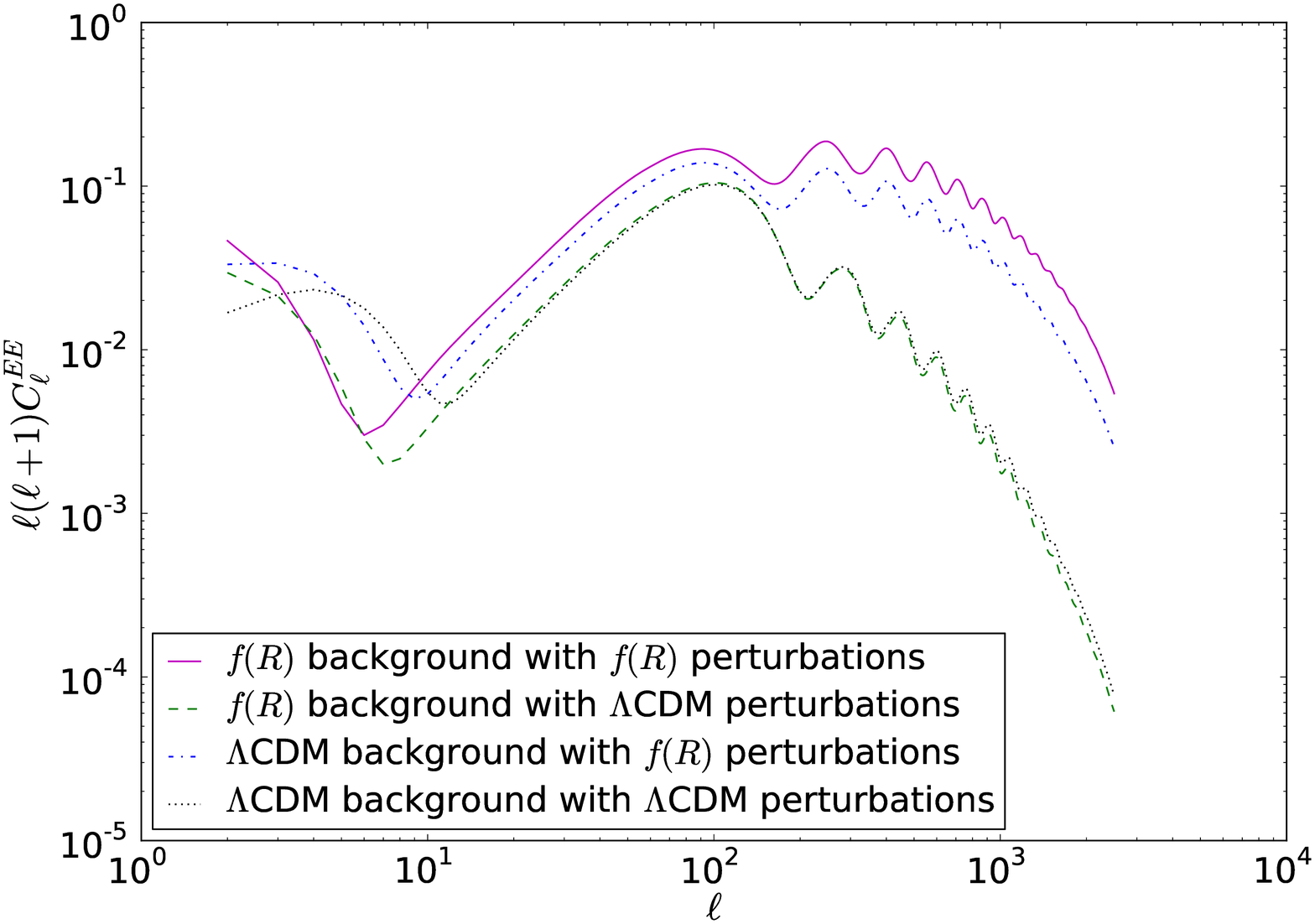}
\\
\includegraphics[width=0.5\textwidth]{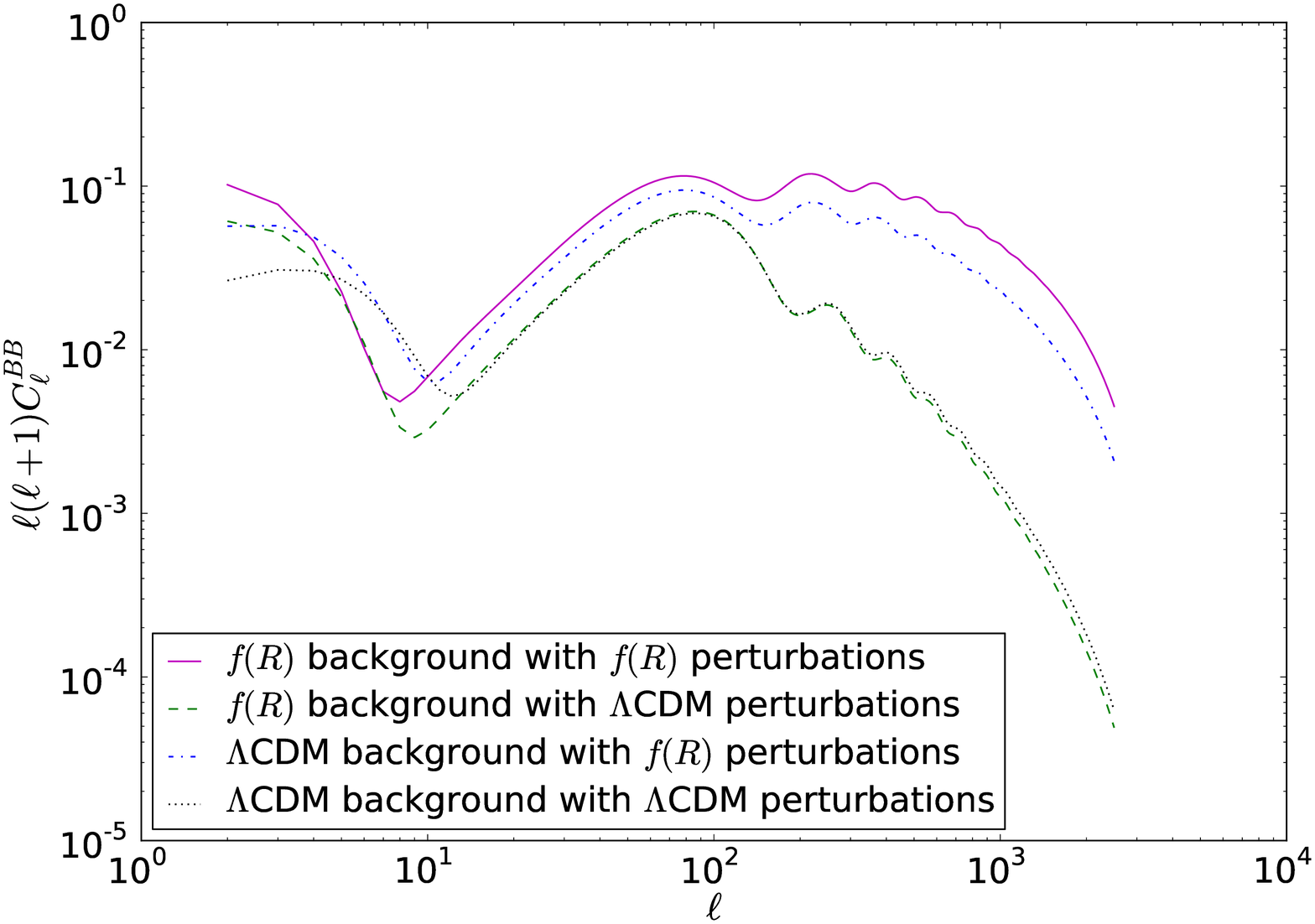}
\caption{
\footnotesize{
Power spectra for the $R^n$ model with $n=1.29$.  The $TT$ (top), $EE$ (centre) and $BB$ (bottom) power spectra for tensor perturbations in all the possible background and perturbations scenarios. Note the similarity between the case of $f(R)$ background with $\Lambda$CDM perturbations and that of $\Lambda$CDM background with $\Lambda$CDM perturbations. 
}}
 \label{n_129}
\end{figure}

\begin{figure}[htb!]
\includegraphics[width=0.5\textwidth]{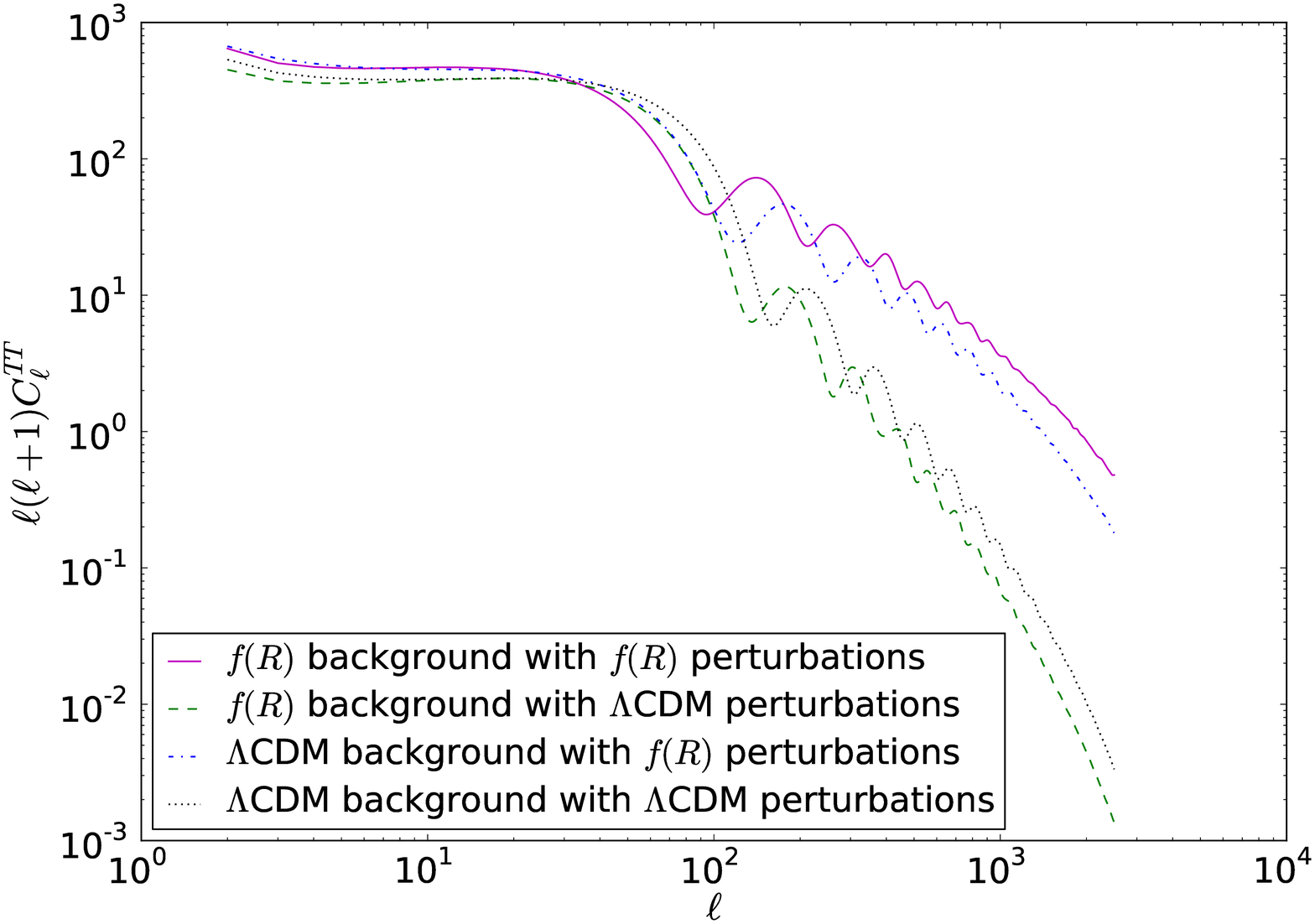}
\\
\includegraphics[width=0.5\textwidth]{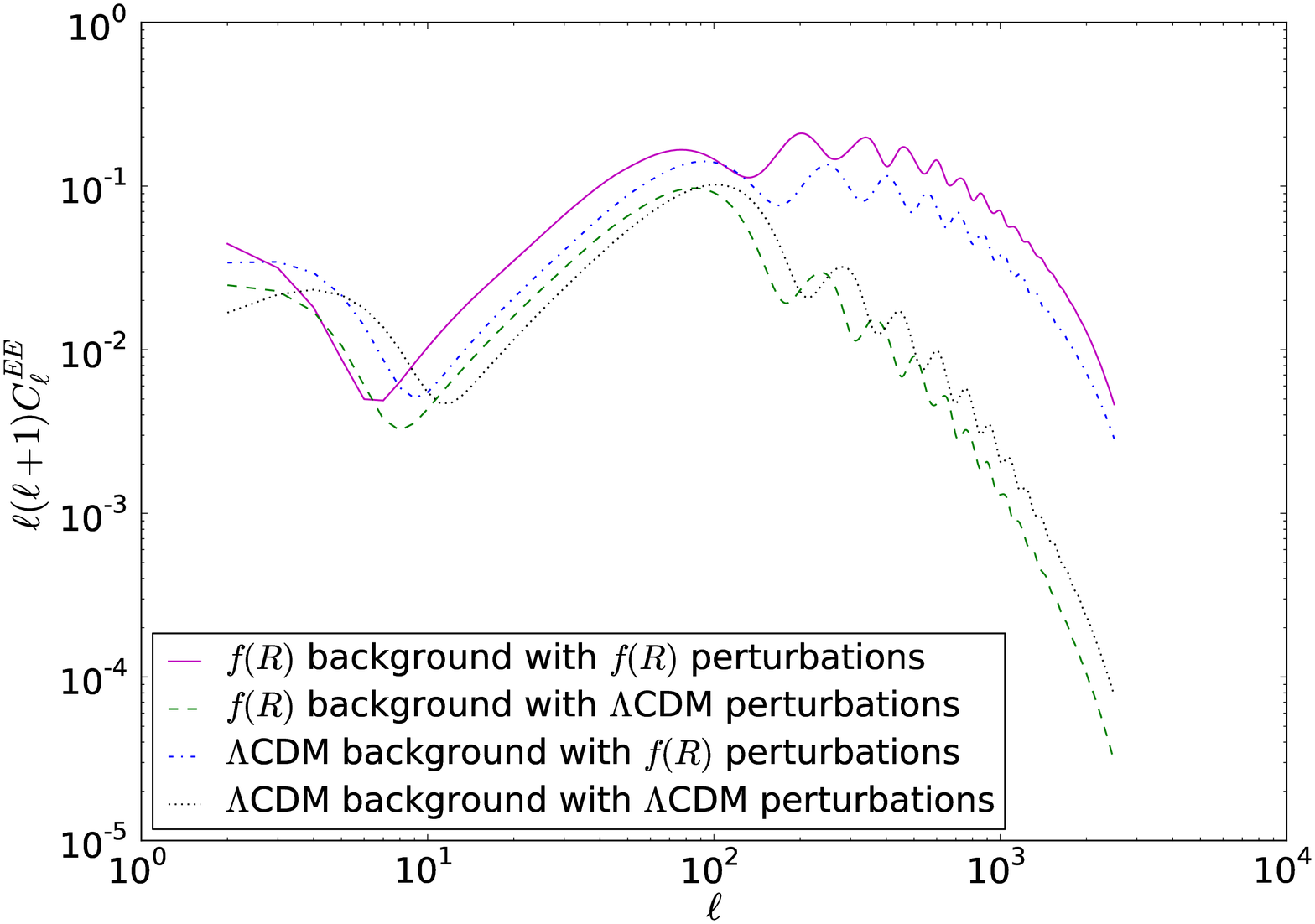}
\\
\includegraphics[width=0.5\textwidth]{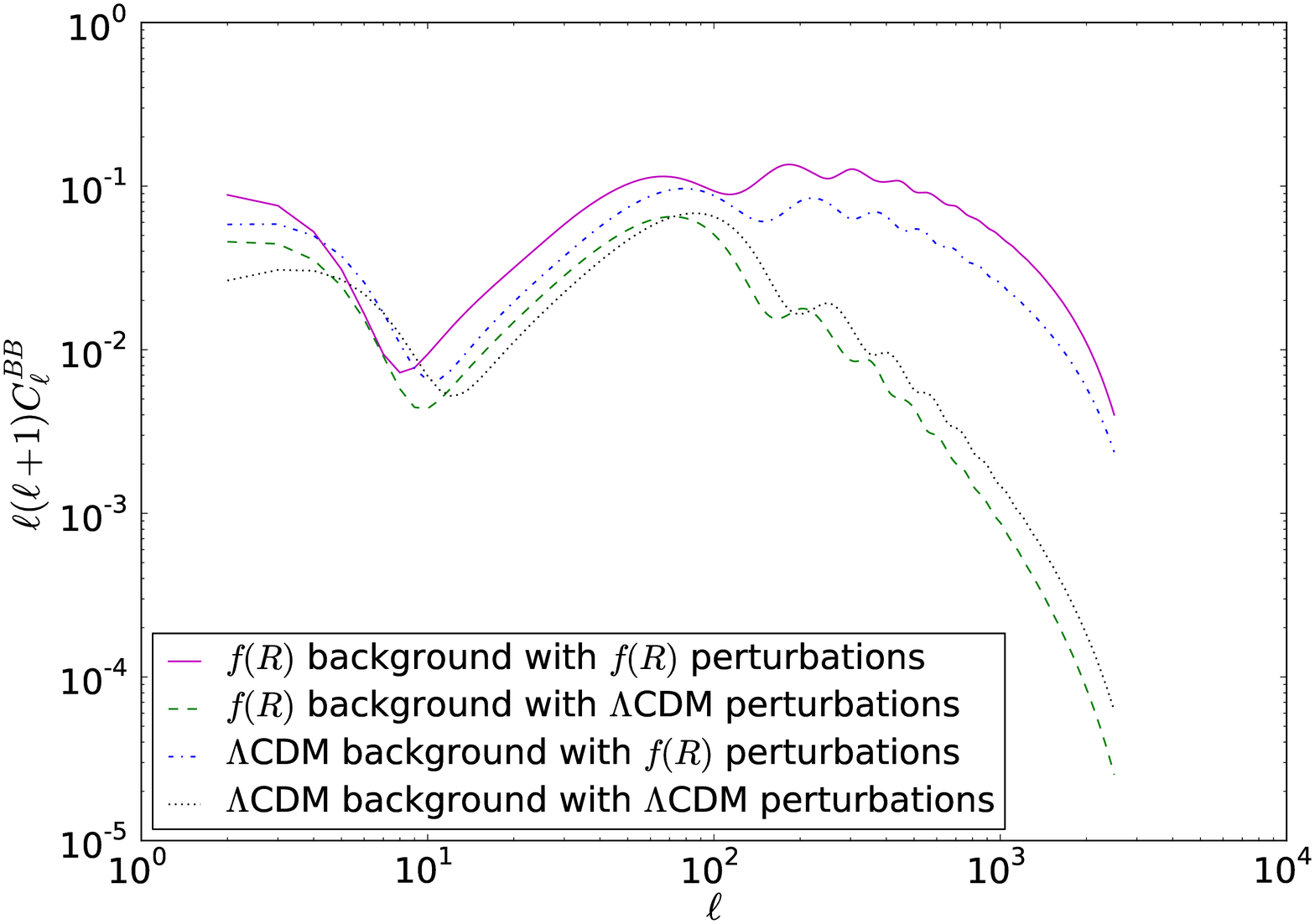}
\caption{
\footnotesize{
Power spectra for the $R^n$ model with $n=1.30$. The $TT$ (top) and $EE$ (centre) and $BB$ (bottom) power spectra for tensor perturbations in all the possible background and perturbations scenarios. Note the distinction between the case of $f(R)$ background with $\Lambda$CDM perturbations and that of $\Lambda$CDM background with $\Lambda$CDM perturbations. Contrast results presented here with those in Figure \ref{n_129}.}}
 \label{n_130}
\end{figure}

In the following, we are interested in features that arise from modifying the background dynamics and those emanating from modifying the perturbation equations. To this end, we compute the power spectra for the four following combinations of background and perturbations, namely 
\begin{itemize}
\item[{\bf a)}] $f(R)$ background with $f(R)$ perturbations, 
\item[{\bf b)}] $f(R)$ background with $\Lambda$CDM perturbations, 
\item[{\bf c)}] $\Lambda$CDM background with $f(R)$ perturbations, and 
\item[{\bf d)}] $\Lambda$CDM background with $\Lambda$CDM perturbations. 
\end{itemize}
For $f(R)$ able to mimic the evolution history of $\Lambda$CDM models, we expect variations in the main features of the power spectra, with respect to $\Lambda$CDM results, to arise from modifications of the perturbation equations. This is the case for the $n=1.29$ model. We note from Figure \ref{n_129} that cases {\bf b)} and {\bf d)} above are almost indistinguishable, owing to the fact that the $n=1.29$ model resembles the expansion history of the $\Lambda$CDM background as shown in \cite{Amare_PRD2013}.
However, this is not a generic feature of $f(R)$ models since most of them usually differ from $\Lambda$CDM at the background level. For these models it turns out to be crucial that one implements the $f(R)$ perturbations on the corresponding $f(R)$ background. For example, in the case $n=1.30$, the spectra for cases {\bf b)} and {\bf d)} are now clearly distinguishable as can be seen in Figure \ref{n_130}.

\begin{figure}[htb!]
\includegraphics[width=0.5\textwidth]{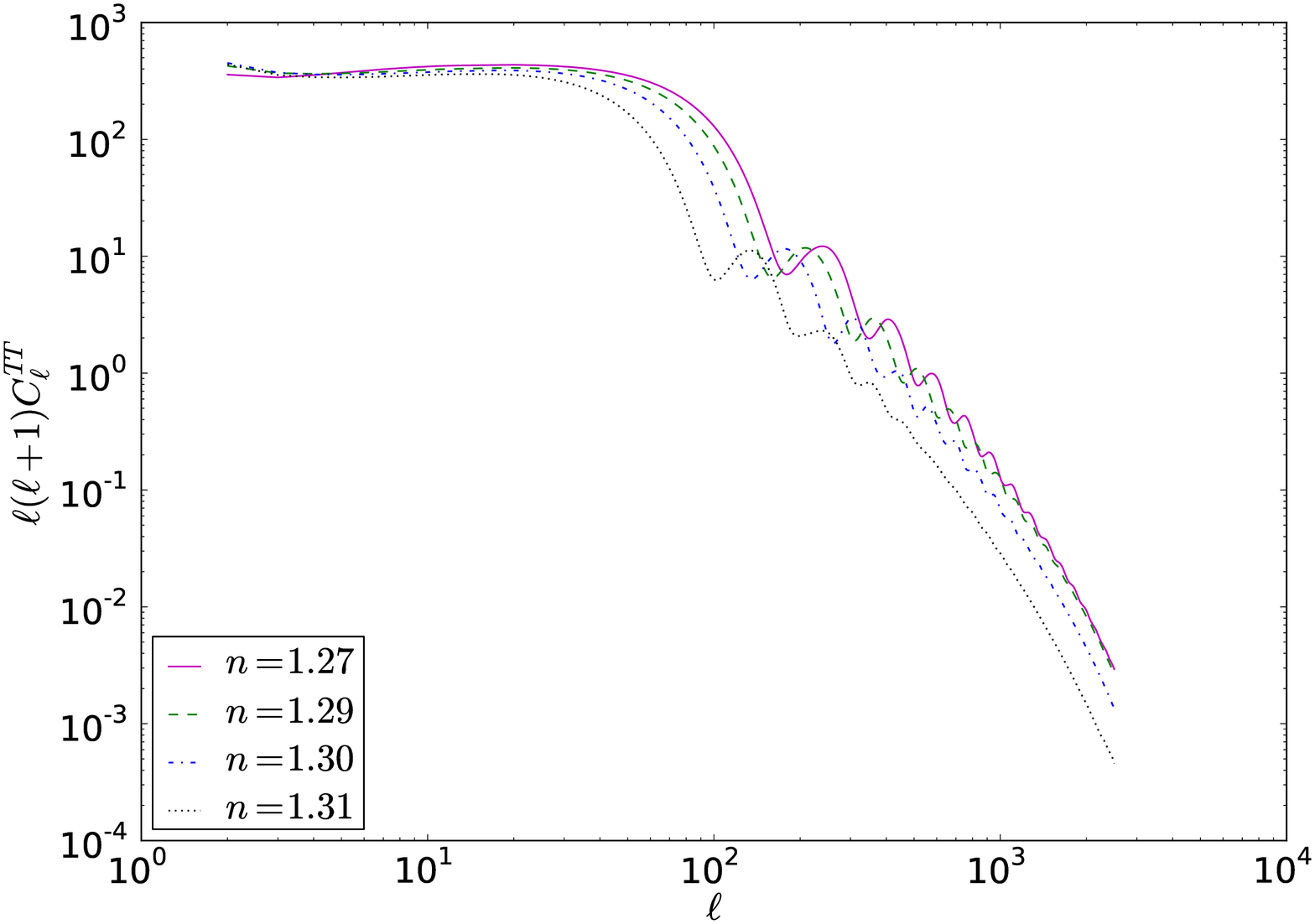}
\\
\includegraphics[width=0.5\textwidth]{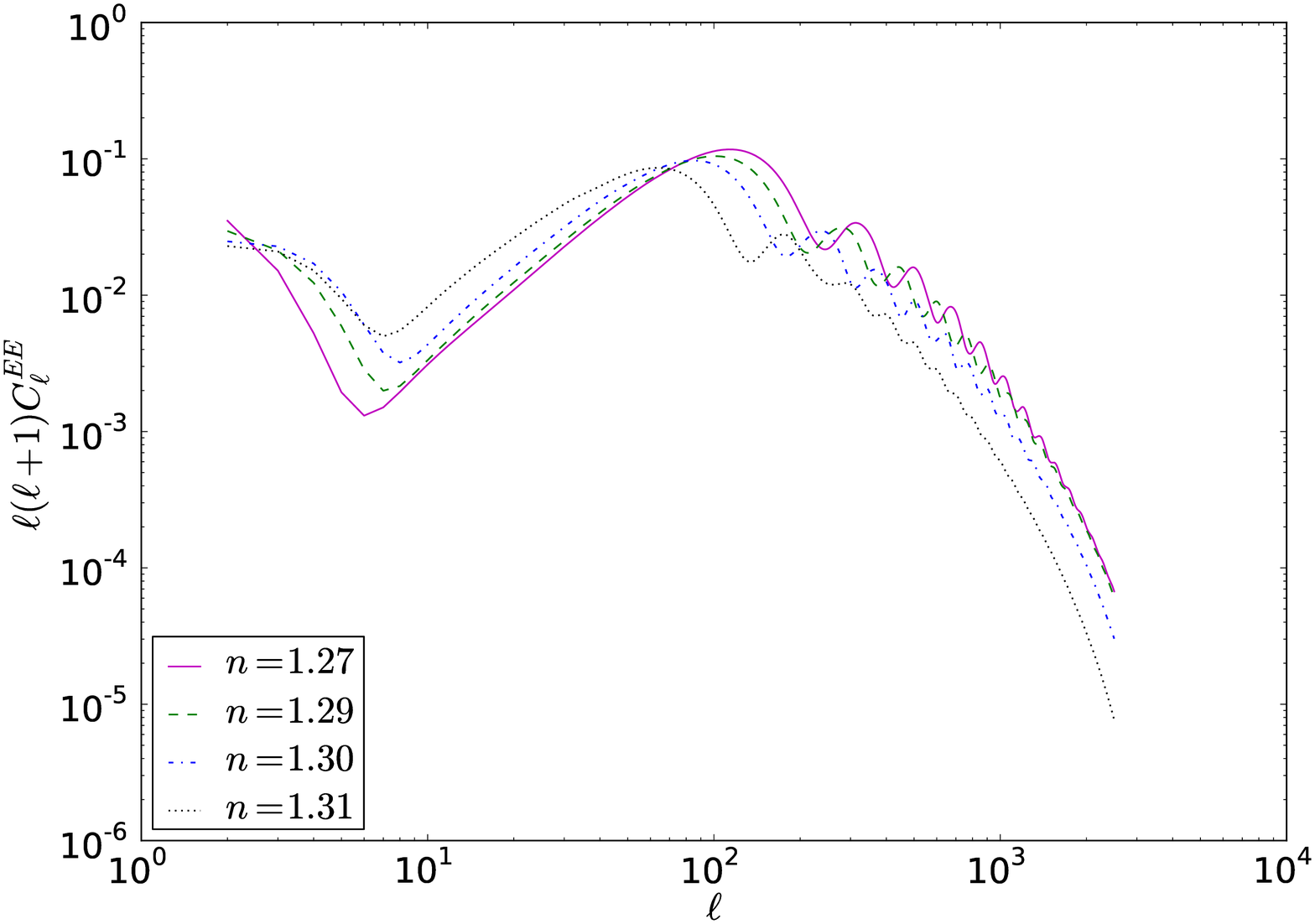}
\\
\includegraphics[width=0.5\textwidth]{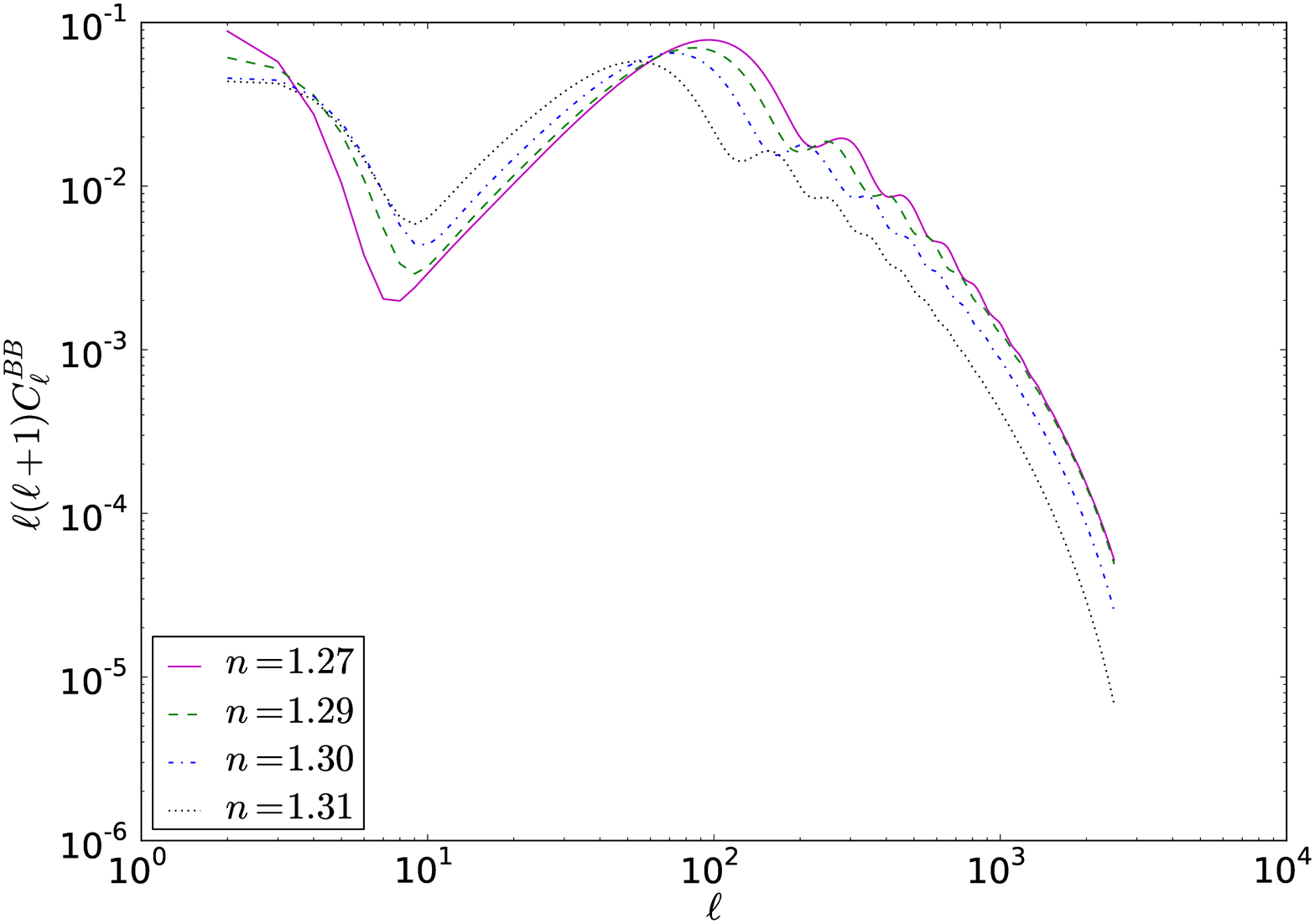}
\caption{\footnotesize{
This figure shows the $TT$ (top), $EE$ (centre) and $BB$ (bottom) power spectra for $\Lambda$CDM tensor perturbations on the $R^n$ model background with $n=\{1.27,1.29,1.30,1.31\}$. 
}}
\label{frgr_all_n}
\end{figure}

More insight on the role of background dynamics and the $f(R)$ corrections to the perturbations can be obtained by studying the combinations {\bf b)} and {\bf c)} above as a function of the parameter $n$. To study the effect of background dynamics, we look at the evolution of $\Lambda$CDM perturbations on an $f(R)$ background for the different values of $n$. This is illustrated in Figure \ref{frgr_all_n}. We observe a drop in power by almost an order of magnitude from $n=1.27$ to $n=1.31$ for both the $TT$, $EE$ and $BB$ spectra. As expected, the effect of the $f(R)$ background is to suppress power on large $\ell$. This effect can be attributed to the damping of modes by the background expansion. It is noteworthy that in the models considered here, the Hubble parameter increases more rapidly with increasing values of $n$. Accordingly, the suppression increases with increasing values of the parameter $n$. This phenomenon clearly demonstrates the need for the correct background model to be implemented in studying anisotropies in modified gravity theories.

\begin{figure}[htb!]
\includegraphics[width=0.5\textwidth]{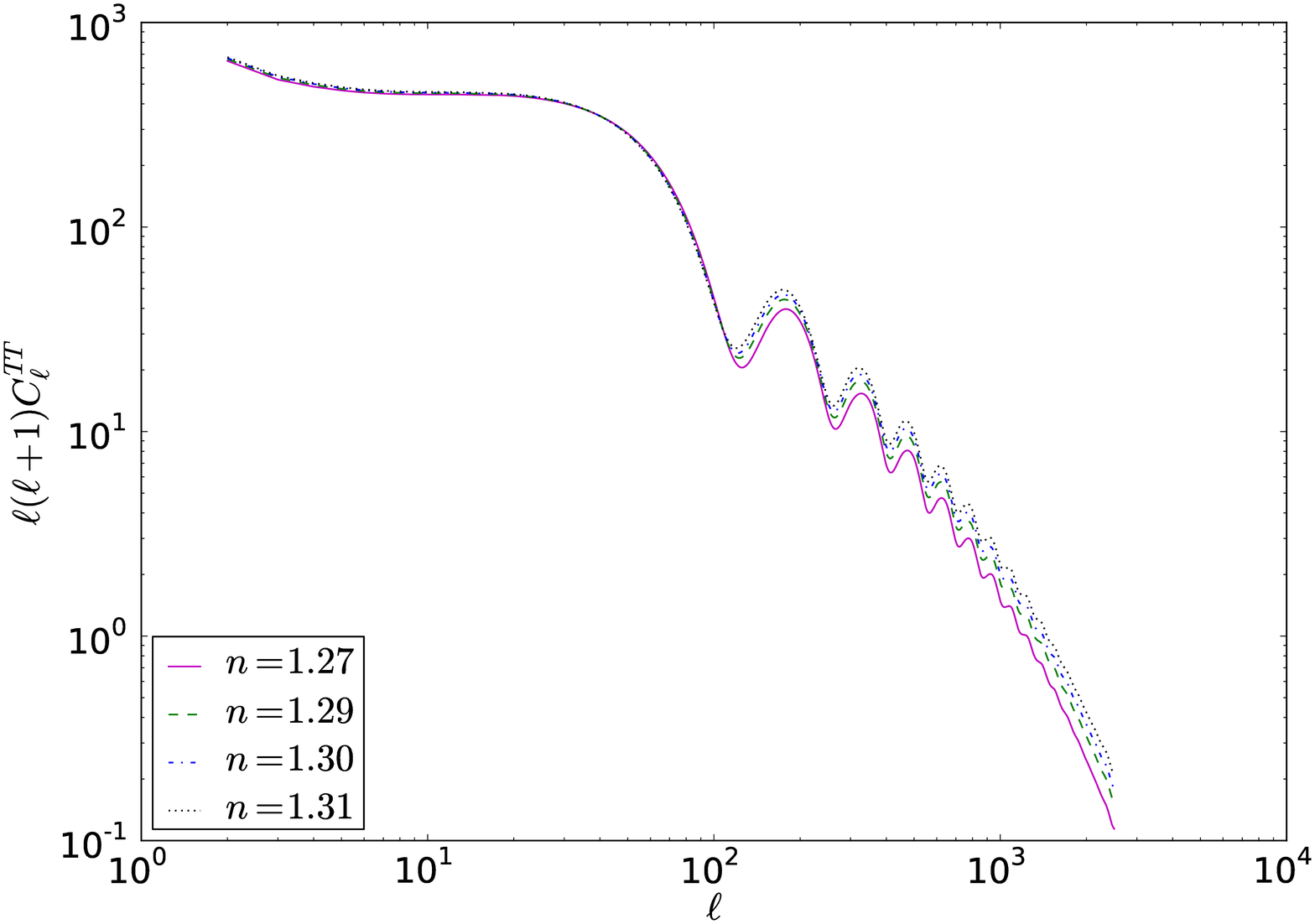}
\\
\includegraphics[width=0.5\textwidth]{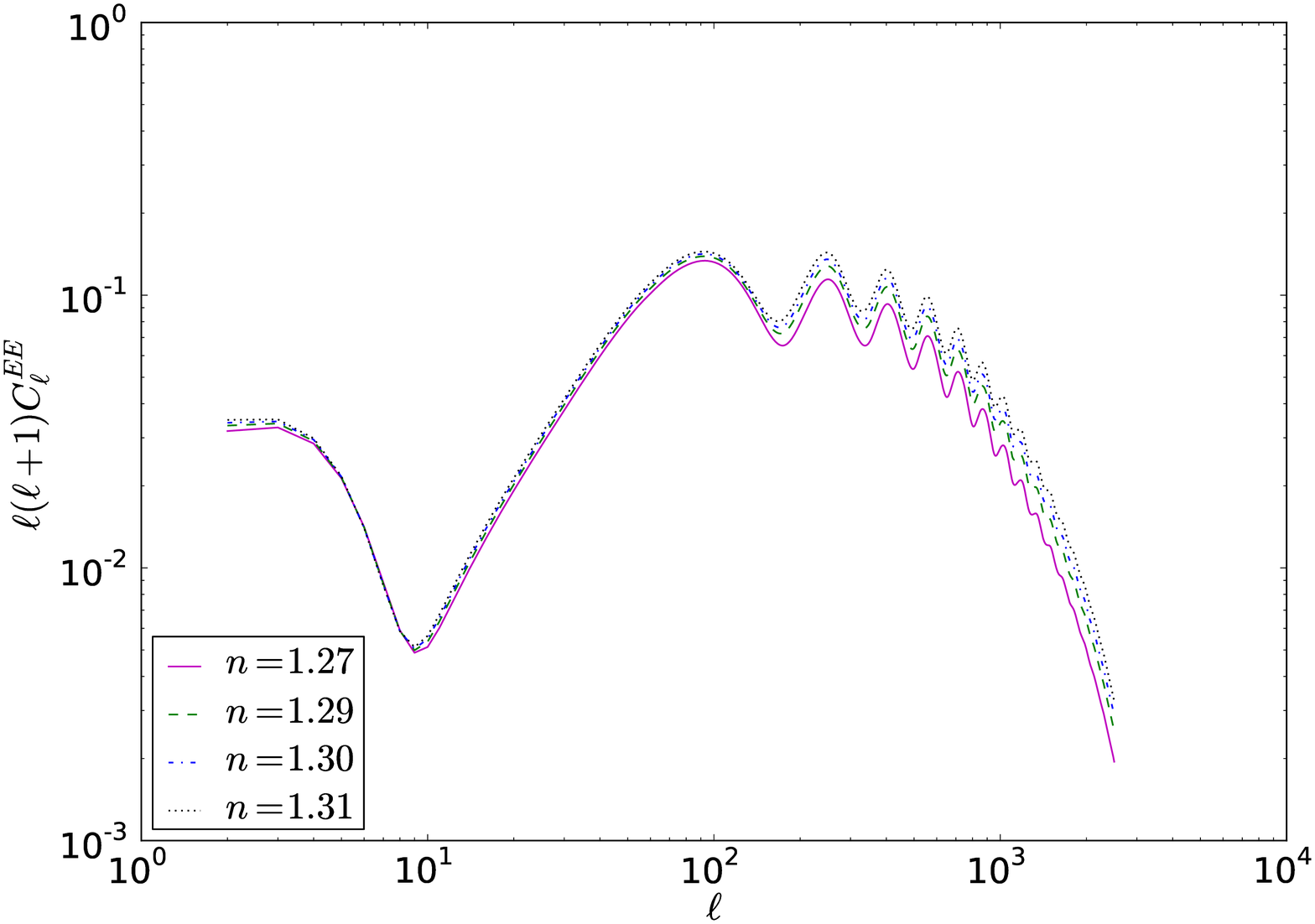}
\\
\includegraphics[width=0.5\textwidth]{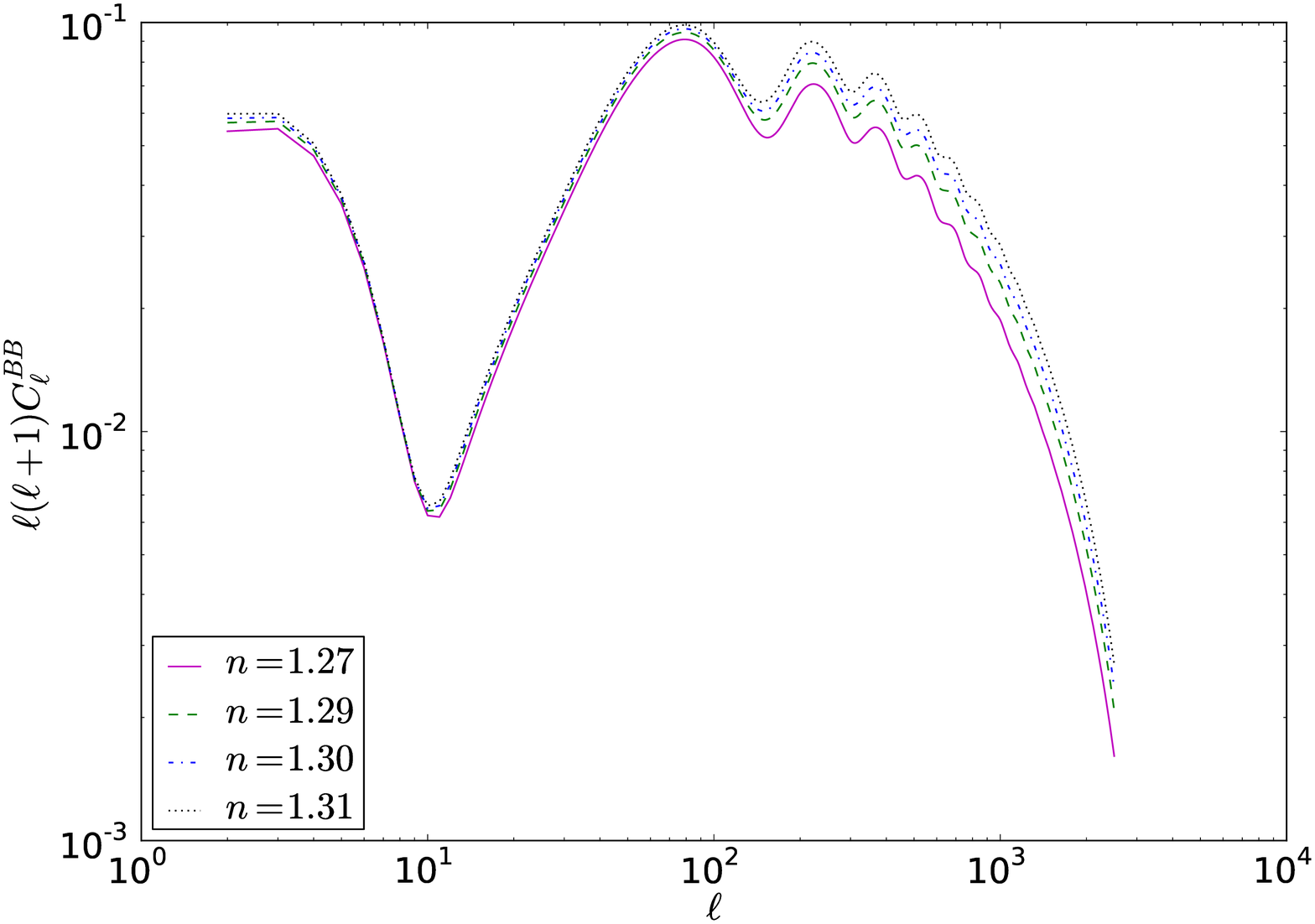}
 \caption{\footnotesize{
This figure shows the $TT$ (top), $EE$ (centre) and $BB$ (bottom) power spectra for $f(R)$ tensor perturbations on the $\Lambda$CDM background for $n=\{1.27,1.29,1.30,1.31\}$. 
}}
\label{grfr_all_n}
\end{figure}

The effects of $f(R)$ corrections on the perturbations can be isolated by considering the combination of a $\Lambda$CDM background with $f(R)$ perturbations. This is shown in Figure \ref{grfr_all_n}. Unlike in the background case, the $f(R)$ contributions to the tensor anisotropies act in a way to boost power at large $\ell$. This phenomenon is not surprising, considering the form of the curvature fluid contribution to the tensor anisotropies. We note from \eqref{pi} that the $f(R)$ corrections to the tensor anisotropies depend on the value of $n$ through the factor $n-1$ and also through the ratio $\dot{R}/R$. When considering $f(R)$ perturbations on a $\Lambda$CDM background, the ratio $\dot{R}/R$ remains the same for all $f(R)$ models and is independent of $n$. In this case, the $f(R)$ corrections are proportional only to $n-1$ and become more pronounced with increasing values of $n$. Interestingly, the features in the power spectra coming from the $f(R)$ perturbations -- on a $\Lambda$CDM background --  are not as pronounced as those coming from the $f(R)$ background -- with $\Lambda$CDM perturbations -- for both the $TT$, $EE$ and $BB$ spectra. As a result, one is likely to underestimate the overall effects of $f(R)$ corrections when considering a $\Lambda$CDM background.

In addition to issues of power amplitude, there is an overall negative phase shift in the oscillations of the power spectra at sub-horizon scales. This aspect of the $f(R)$ corrections is most notable in Figure \ref{frgr_all_n} for all the studied spectra. Since Figure \ref{frgr_all_n} considers $f(R)$ backgrounds, we attribute the phase shift to background dynamics. This is consistent with the fact that for low $\ell$, a similar shift is observed when an $f(R)$ background is employed for the $EE$ and $BB$ spectra in Figures \ref{n_129} and \ref{n_130}.

\section{Conclusions}
\label{Conclusions}

In this work we have presented for the first time a detailed analysis of the tensor CMB features for a simple one parameter class of $f(R)$ modified gravity theories by using a new implementation of CAMB. These simulations used both the correct cosmological background evolution as provided by these fourth-order gravity theories as well as the required tensor perturbation equations. 

We applied our general results to $R^n$ models for different values of $n$ verifying the convergence to General Relativity result when $n$ 
approaches unity and describing the features that may distinguish those models from Concordance $\Lambda$CDM model predictions. Our implementation demonstrates the importance of considering the correct background when alternative theories of gravity are subjected to this kind of analysis.

According to our results, by only considering perturbations for the $c_{\ell}^{TT}$, having assumed the usual General Relativity background expansion history, would lead for instance to not seeing any appreciable difference between pure General Relativity and the $n=1.29$ cases for all the depicted spectra. This fact can be manifestly seen in the top panel of Figure \ref{n_129}. Moreover, values in the interval $n=1.27-1.31$ are also indistinguishable from each other at low $\ell$'s, although for high $\ell$'s this degeneracy is broken as can be seen in the top panel Figure \ref{grfr_all_n}. This stresses the need of a full consideration of the correct background. With regards to the $c_{\ell}^{EE}$ and $c_{\ell}^{BB}$  coefficients, it can be seen for instance how for $n=1.29$ the sole consideration of perturbations (keeping the $\Lambda$CDM background) is hardly detectable for intermediate $\ell$'s, although differences appear at low and high $\ell$'s. This fact can be manifestly seen in the centre and bottom panels of Figure \ref{n_129}. 

Our code therefore provides a powerful tool capable of determining the key features of the effect that fourth-order gravity theories have on CMB tensor perturbations, when a specific $f(R)$ model is claimed as viable. This will be presented in a future work for other models available in the literature.
In this realm, it is also necessary to investigate the impact of scalar perturbations for models under consideration, which have a dominant contribution to the measured power spectra. Exclusion tests for $f(R)$ models can be performed since data for $c_{\ell}^{TT}$ are available from WMAP \cite{WMAP} and more recently from Planck \cite{Planck} once the scalar contribution is included. 
With respect to the polarisation anisotropies, Planck data (and its joint analysis with BICEP2) may allow us to constrain the $E$ and $B$-modes in the near future. One particularly interesting feature of tensor modes in these theories is 
the appearance of an amplification in the $B$-modes power spectrum, in particular around the peak at $\ell \approx 80$, due to the different evolution of tensor perturbations. Figures  (\ref{n_129}) and (\ref{n_130}) may help to visualise this statement. This enhancement would clearly favour the detectability of primordial gravitational waves in these models compared to General Relativity. Consequently inflationary models occurring on lower energy scales when combined with the models under consideration in this investigation would present the same 
$B$-mode amplitude as if the tensor perturbations had been evolving as in $\Lambda$CDM model for commonly accepted inflationary scales. This means that these theories present a degeneracy between the model parameters and the inflationary scale. In other words, the assumption that the detection of $B$-modes directly fixes the scale of inflation should be cautiously reconsidered until one better understands how combined effects coming from gravitational theories beyond General Relativity, together with the modification of the inflationary scale may have on the amplitude of the $B$-modes. 

The tests described in this paper can be applied in order to further constrain 
the allowed range of parameters for classes of $f(R)$ models that provide an alternative explanation for the late-time cosmological speed-up. In particular, for models mimicking General Relativity at high redshift, the formalism presented here can be directly applied, since Equation (\ref{IC2}) will be immediately satisfied. Work to include scalar anisotropies in CAMB in order to compare with present and future data is in progress.

\begin{acknowledgements}
We would like to thank Garry Agnus and Fernando Atrio  for useful help about CAMB technicalities and Jos\'e Beltr\'an and Antonio L. Maroto for helpful discussions. 
We also thank CosmoCoffee forum website.
A.d.l.C.D. acknowledges financial support from MINECO (Spain) projects FPA2011-27853-C02-01, FIS2011-23000 and Consolider-Ingenio MULTIDARK CSD2009-00064.
A.d.l.C.D. is indebted to the Astrophysics, Cosmology and Gravity Centre (ACGC), Department of Mathematics and Applied Mathematics, University of Cape Town, South Africa for its hospitality during the latest stages of this manuscript preparation. 
P. K. S. D. thanks the NRF for financial support.\\
\end{acknowledgements}

\appendix
\section{CAMB modifications}
\label{sec:appendix}
For a complete specification of the background model, we need values for $w_{eff}$, $f_{R}$, $\dot{R}/R$ specifying the expansion history in terms of the scale factor $a$.
These specifications are given through data files once the dynamical system (\ref{DS2}) has been integrated.
We give the values at logarithmically spaced intervals of the scale factor $a$, ranging from radiation dominated epoch till today 
as required in the CAMB computations.
Any necessary interpolations along $a$ are carried out using the $\texttt{spline}$ subroutines that are built within CAMB.
For the effective equation of state parameter $w_{eff}(a)$, the procedures required in order to read $w_{eff}(a)$ from a data file have already been implemented within CAMB. For our modifications, we edit the file \texttt{equations.f90} to add subroutines $\texttt{get\_ricci(a)}$ and $\texttt{one\_over\_fprime(a)}$ which read and store values for $\dot{R}/R$ and $1/f_{R}$ respectively from the user-supplied data files. These routines can then be used to supply values for $\dot{R}/R$ and $1/f_{R}$ at scale factor values in the desired range. 
In general, the background modifications involve inclusion of the $1/f_{R}$ factor in the matter variables that occur in the CAMB functions \texttt{dtauda()} and \texttt{fderivst()}, both of which are found in the \texttt{equations.f90} file. 

For the perturbation equation, we modify \texttt{fderivst()} to take into account $f(R)$ corrections to the tensor anisotropies. The relevant variables in this case are \texttt{aytprime(2)} and \texttt{aytprime(3)} which give the time evolution of the tensor part of the Bardeen variable $H_{\scriptscriptstyle T}^{(\scriptscriptstyle 2)}$ and the shear $\sigma_{(k)}$ respectively. In the metric approach to perturbation theory, tensor perturbations are encoded in the variable $H_{\scriptscriptstyle T}^{(\scriptscriptstyle 2)}$. For $f(R)$ models, this can be shown to obey the wave equation,
\begin{align}
\label{eq:tensor_bardeen}
& H_{\scriptscriptstyle T}^{\scriptscriptstyle (2)}{}''_{}+\left[2\mathcal{H} + (n-1)\frac{R'}{R}\right]H_{\scriptscriptstyle T}^{\scriptscriptstyle (2)}{}' \nonumber \\
&+(2K+k^{2})H_{\scriptscriptstyle T}^{\scriptscriptstyle (2)}{}\,=\, \frac{p^{(m)}a^{2}\pi^{(m)}_{k}}{f_{R}}\;,
\end{align}
where $K$ holds for the 3-curvature of spatial sections and $\pi^{(m)}$ the matter anisotropic stress.
 One can rewrite \eqref{eq:tensor_bardeen} as a first order differential equation for the shear $\sigma$ by using the identification $H_{\scriptscriptstyle T}^{\scriptscriptstyle (2)}{}'_{}=-k\sigma_{(k)}$, corresponding to \texttt{aytprime(2)} in CAMB. The shear evolution then becomes,
\begin{align}
 \sigma'_{(k)}+ \left[2\mathcal{H} + (n-1)\frac{R'}{R}\right]\sigma_{(k)} 
-k H_{\scriptscriptstyle T}^{(\scriptscriptstyle 2)}
=- \frac{p^{(m)}a^2\pi^{(m)}_{k}}{k f_{R}}\;,
\end{align}
where the two equations above are expressed in geometrised units again ($8\pi G\equiv1$) and we have set $K=0$ in the last equation. 
Some auxiliary modifications of other CAMB modules were necessary. For example, in order to control the convergence of integrations within the \texttt{rombint2()} function (found in \texttt{subroutines.f90}), we have had to increase the number of maximum iterations MAXITER. Moreover, since our models are spatially flat, one has to be careful that the input cosmological parameters are consistent with flat geometry to the required accuracy as given by $\texttt{OmegaKFlat = 5e-7}$ in \texttt{modules.f90}. We also edit \texttt{inidriver.f90} to be able to read the value of $n$ corresponding to the $f(R)$ function from the parameter file \texttt{params.ini}.


\end{document}